\documentstyle[11pt,fleqn]{article}
\def\ep {\epsilon}

\def\e2 {\epsilon-\epsilon_k}
\def\be {\begin{equation}}
\def\ee {\end{equation}}
\def\bea {\begin{eqnarray}}
\def\eea {\end{eqnarray}}
\def\om {\omega}

\begin{document} 
\begin{small}
\begin{center}
{\bf Paramagnons, weak disorder and positive giant magnetoresistance }
\mbox{}

{George Kastrinakis}

{\em Department of Physics, University of Illinois at Urbana-Champaign,
Urbana, IL 61801-3080}

{(Oct. 25, 1997)}

\end{center}

\begin{abstract}

At low temperature and for finite spin scattering in a
weakly disordered metal, for a certain value, predicted from our theory,
of the material-dependent
paramagnon interaction, the total conductivity becomes highly sensitive to
the orbital effects of a finite magnetic field.
As a consequence, positive giant magnetoresistance
and giant corrections to the Hall coefficient arise.
We obtain very good agreement between this theory
and recent positive giant magnetoresistance experiments, while
making specific material-dependent predictions.
                 
\end{abstract}

\mbox{}

	Recently, there has been a plethora of both experimental and 
theoretical investigations of giant magnetoresistance (GMR)
in metallic systems \cite{xi}. In most cases experimentally observed
so far, increasing the magnetic field $H$ from zero causes the
resistance to decrease to a fraction of its zero field value. 
This behavior persists for temperatures ranging from zero to 
well above room temperature.
However, in the experiment of Tsui, Uher and Flynn\cite{tuf} the observed
GMR differs {\em drastically} from the usually observed GMR
in four ways. 1) The effect only exists at {\em low} temperature $T$,
with the magnetoresistance correction reaching $\sim$ 35\% of the zero
field value for $H\sim 6$T, but vanishing completely
above $60^o$K for $H\leq6$T. 
2) GMR is {\em anisotropic} with regards to the direction of the field $H$, 
3) it is {\em not} connected to the magnetization and does {\em not} 
saturate with increasing $H$, for fields as big as at least 8T, and 
4) it is positive, i.e. it increases with $H$.
Save for the giant magnitude of the effect, the four characteristics above
can be explained in the frame of the metallic weakly
disordered regime ${\epsilon_F \tau}\gg 1$, where $\epsilon_F$ is the Fermi 
energy and $\tau$ the elastic scattering time arising from disorder
\cite{lr,aa}. In this 
regime, the conductivity corrections, due to disorder induced diffusion
and to electron-electron interactions, are of order $\sigma_o/
({\epsilon_F \tau})^r$, where $\sigma_o$ is the Drude term and $r=1,2$ 
for $d=2,3$ space dimensions.

We thereby propose a novel mechanism for {\em giant} 
corrections to the transport quantities, including GMR, due to the presence 
of paramagnons in a {\em weakly disordered} metal.
At {\em low} temperature and for {\em finite} impurity {\em spin} scattering, 
for a certain value, predicted from our theory, of the material-dependent
paramagnon interaction, the total conductivity becomes highly sensitive to 
the orbital effects \cite{lr} of a finite magnetic field. 
This is attributed to certain 
microscopic processes, otherwise negligibly small, which can be enhanced
by a resonance factor,
emanating from the 
spin-density channel.
Thus an experimental 
signature like the one observed by Tsui et al. \cite{tuf} is obtained. 
As we explain below, the samples used in ref. \cite{tuf} contain
the ingredients necessary for the appearance of GMR, in accordance with our
theory.
 
	We begin by considering a constant paramagnon interaction $A_o$
acting {\em only} between particles 
of opposite spin and given by
\begin{equation}
A_o = \Phi/N_F \;\; ,
\end{equation}
where $\Phi$ is dimensionless and positive, and $N_F$ is the density of states
at the Fermi level. 

	In the presence of weak disorder, which includes {\em spin} scattering,
the ladder diagrams in the particle-hole channel 
give rise to a propagator $A^j(q,\omega)$. $j=-1,0,1$ is the total spin
difference between particle and hole of these spin-density propagators.
$A^j$ obey the coupled Bethe-Salpeter equations
\begin{equation}
A^1=A_o + A_o{\cal D}^1 A^1 + A_o {\cal D}^0 A^0 \;\;,  \label{ex1}
\end{equation}
\begin{equation}
A^0= A_o{\cal D}^0 A^1 + A_o {\cal D}^{-1} A^0 \;\;,	\label{ex2}
\end{equation}
shown in fig. 1 - here we supressed the variables $q,\omega$, which stand
for the momentum and energy difference between particle and hole lines,
respectively. Note the explicit spin indices in the figure.
${\cal D}^j$ are given by 
the components of the density and spin-density correlation functions
\cite{aaz,ml}
${ \cal D}^{\pm1}={\cal D}^{1,\pm1},\; 
{\cal D}^0=
[ {\cal D}^{0,0} - {\cal D}^{1,0}]/2,\;
{\cal D}^{j,m}(q,\omega)=N_F 
\{Dq^2+j4\tau_{S}^{-1}/3\}/ \{Dq^2+j4\tau_{S}^{-1}/3-i\omega-im\omega_H \}
$, with $\tau_S^{-1}$ being the total spin scattering rate ($\hbar=c=1$), 
$D$ the diffusion constant and $\omega_H$ the Zeeman energy.
$A^{-1}$ obeys the same set of coupled equations with all spin indices
reversed - equivalently $A^{-1}(\omega_H)=A^{1}(-\omega_H)$.
The solution of these equations is written in the form
\begin{equation}
A^j(q,\omega) = \frac{K_{\Phi j} Dq^2 - i L_{\Phi j} \omega +M_{\Phi j}}
{A_\Phi Dq^2 - iB_\Phi \omega + C_\Phi } \;\;,
\end{equation}
with the coefficients $A_\Phi,B_\Phi,C_\Phi,K_{\Phi j},L_{\Phi j},M_{\Phi j}$ 
{\em depending} on $\Phi$ and $H$.
There are {\em two} limiting cases for $A_\Phi,B_\Phi,C_\Phi$ 
as a result of the finite
spin scattering rate $\tau_S^{-1}$. For $Dq^2 > \omega$ 
("static limit") we have
\bea
A_\Phi=4(1-\Phi)^2 \;\; , \;\; B_\Phi=12-22\Phi+10\Phi^2+O(\Omega_{H}^2)\;\;,\\ 
C_\Phi=[\;(1-\Phi)^2+O(\Omega_{H}^2)\;]\; 4\tau_S^{-1}/3 \;\;, \nonumber
\eea
while for $Dq^2 < \omega$ ("dynamic limit") we have 
\bea
A_\Phi=12-20\Phi+15\Phi^2/2+O(\Omega_{H}^2) \;\;, \;\; 
B_\Phi=4-6\Phi+3\Phi^2/2+O(\Omega_{H}^2)\;\;, \\  \label{dyna}
C_\Phi=[\;1-2\Phi+3\Phi^2/4+O(\Omega_{H}^2)\;] 4\tau_S^{-1}/3 \;\;, \nonumber
\eea
where $\Omega_H = \omega_H \tau_S$.
It is explicitly assumed that $\Phi$ does not approach 1 closely,
which would be the onset of the ferromagnetic transition.

	The effect of giant magnetoresistance is attributed (see also below)
to the combination of a negative $C_\Phi$, for $\Phi = 2/3 - 2$,
{\em and} a vanishing $B_\Phi$, for $\Phi \simeq \Phi_o \equiv 0.845$, in the
{\em dynamic limit}. These two conditions can
yield a resonance in the denominator of the propagator $A(q,\om)$,
which results in the enhancement of certain diagrammatic processes,
otherwise negligibly small, which have a very high sensitivity to the
presence of a magnetic field, thus causing the appearance of GMR.
We will see below that there are materials satisfying $\Phi \simeq \Phi_o$.

	We observe that diagrams involving a factor $Z_m = \sum_\om 
\int d\vec q \; A^m(q,\om) \; F(q,\om)$ can yield a {\em large}
contribution {\em if} $m=2 n$. This is the case because $Z_m = \sum_\om 
\; z_m(\om)$, with 
\begin{equation}
z_{2n} \simeq F \int_{x_o-a}^{x_o+a} \frac{dx}{(x-x_o- i \delta)^{2n}}
\simeq \frac{2\; F}{(2n-1) \; \delta^{2n-1}} \;\;.
\end{equation}
Here $\delta \equiv B_\Phi\; \om \rightarrow 0, \; x \equiv A_\Phi Dq^2, 
\; x_o \equiv - C_\Phi>0, \; a < x_o$ and $F \equiv F(x=x_o,\om)$
(finally we will only make use of the case $n=1$). 
After appropriate $\om$ Matsubara summation, a term $B_\Phi^{2n-1}$ remains
in the denominator 
yielding an overall enhancement factor.

	The relevant dominant class of diagrams containing factors $A^{2}(q,\om)$
are shown in fig. 2. These diagrams can be sandwiched between two pairs
of $G^R(k,\epsilon) G^A(k,\epsilon)$ to yield conductivity contributions.
Let us first take a close look at fig. 2a.
The 2 Cooperons are inserted between the 2 $A(q,\om)$'s for the 
following reasons. 1) They introduce a magnetic field dependence of the
conductivity, which will be in accordance with the experiments of ref. 
\cite{tuf} as to the magnitude and direction of the magnetic field etc.
2) If they were absent, the 2 $A(q,\om)$'s would just
collapse onto a single $A(q,\om)$. 3) An {\em even} number of them
is needed in order to properly conserve momentum in this diagram. 
4) Since each Cooperon introduces a small factor $1/(\epsilon_F \tau)$,
we keep only diagrams with 2 Cooperons, and not 4,6,8,..., between any 
2 $A$'s. This small prefactor is counterbalanced by the resonance 
of $A^{2n}$ we mentioned above.

In passing, let us note that Finkelstein\cite{fink}, Castellani, Di Castro,
Lee and Ma\cite{cclm}, and Chang and Abrahams\cite{ca}, have shown
that the diffusive correlators (diffuson and Cooperon) retain their
form in the presence of interactions - modulo a renormalization of 
the diffusion coefficients, inelastic scattering rate etc. 
As a result we do not need to consider
explicit interaction contributions in the diffuson and the Cooperon.

	In the following, we restrict ourselves to the low temperature 
limit $T\rightarrow 0$.

	Now we sum 
the diagrams shown
in figs. 2 a,b,c, which give the dominant contribution to the
paramagnon conductivity here (to lowest order in the parameter $b_H$ in 
eq. (\ref{bh}) below). We also take into account the same diagrams
but with the propagator $A$ substituted by the combination $D A D$,
i.e. a propagator $A(q,\om)$ sandwiched between two diffusons $D(q,\om)$. 
This latter
combination has appeared in the works of Altshuler, Aronov, Larkin and 
Khmelnitskii \cite{aalk}, Lee and Ramakrishnan \cite{lr2} and Millis and Lee 
 \cite{ml}, where the usual small
magnetoresistance due to weak disorder and interactions was investigated.
We find that the total contribution of these diagrammatic blocks
is given by the block $\Gamma_H$
\begin{equation}
\Gamma_H(k,\ep) = \Gamma_{1H} - \Gamma_{2H} \; G^R(k,\ep) G^A(k,\ep) \;\;,\;\;
\Gamma_{1H} = \alpha \; \tau^2 \; b_H \;\;, \;\; 
\Gamma_{2H} = \frac{b_H}{2\pi\tau\epsilon_F} \;\;, \\
\end{equation}
Here 
$G^{R,A}(k,\epsilon)=1/\{\epsilon-\epsilon_k\pm i/2\tau\}$,
$c_H = \sum_q C^2(q,\om=0)$ (see eqs. (\ref{coop}),(\ref{coope}) below) and \\
$s= M_{1\Phi}^2 \; [-C_\Phi/(A_\Phi D)]^{d/2-1} / (2 A_\Phi D)$,
\be 
b_H = \frac{s^2 \; t\; c_H } {B_\Phi^2} \;\; ,\label{bh} 
\ee
$t = 5 \pi \; N_F \; (2\tau)^7 / 16$,
$\alpha = 1 - (2\tau)^2 \left(\frac{4}{9 \tau_S^2}+
\frac{1}{3 \tau_S \tau_o}\right)$ and 
$\tau_o^{-1} = \tau^{-1} - \tau_S^{-1}$ is the scattering rate due 
to non-magnetic impurities.

	Then we take the ladder sum of $\Gamma_H$
as follows
\begin{equation}
B(k,\ep) = \Gamma_H(k,\ep) \; + \; \Gamma_H(k,\ep) G^R(k,\ep) G^A(k,\ep) 
B(k,\ep) 
= \frac{\Gamma_H(k,\ep)}{1 - \Gamma_H(k,\ep) G^R(k,\ep) G^A(k,\ep)} \;\; .
\end{equation}

	The total conductivity due to paramagnons
is given by
\begin{equation}
\sigma_P = \frac{2e^2}{m^2} \int d\vec k \;k_x^2 \; 
\{G^R(k,\ep_F) G^A(k,\ep_F)\}^2 \; B(k,\ep_F) = -\sigma_o \;+\; \sigma_c \;\; .
\end{equation}
$\sigma_o$ is the well known Drude term. Finally the total conductivity
is given by
\begin{equation}
\sigma(H) = \sigma_o \; + \; \sigma_P = \sigma_c \;\;,
\end{equation}
\begin{eqnarray}
\sigma_c = \frac{4 N_F e^2 \epsilon_F }{3 m S_H}
 \left\{ \frac{(\Gamma_{1H}- \frac{\Gamma_{2H}}{y_+})}{ \sqrt{\sigma^2 - y_+} }
\arctan\left( \frac{\epsilon_F}{\sqrt{\sigma^2 - y_+}} \right) 
- \; \frac{(\Gamma_{1H}-\frac{\Gamma_{2H}}{y_-}) }{ \sqrt{ \sigma^2 - y_-} }
\arctan\left( \frac{\epsilon_F}{\sqrt{\sigma^2 - y_-}} \right) \right\}
\;\;. \label{sig}
\end{eqnarray}
Here $y_\pm = (\Gamma_{1H} \pm S_H)/2, \; 
S_H=\sqrt{\Gamma_{1H}^2 - 4 \Gamma_{2H}}$, and
$\sigma\equiv 1/2\tau$.
Note that as the temperature $T$ is increased, $\Gamma_H$ decays due to the 
increasing dephasing rate $\tau_\phi(T)^{-1}$ in the denominator 
of the Cooperons.
As a result, the overall magnitude of $\sigma_P$ - and of the 
magnetoconductivity - decays, and we recover 
the usual Drude conductivity. In $d=2$ with $H$ perpendicular to the system
we have
\be
c_H = \sum_q C^2(q,\om=0) = \frac{y}{ H} \sum_{n=0}^\infty
\left ( \frac{1}{a H}  + n + \frac{1}{2} \right )^{-2} \;\;,  \label{coop}
\ee
with $a=4D e\tau_{\phi}(T)$ and $y= 1/\{2 (4\pi)^3 N_F^2 \tau^4 e D\}$.
Moreover
in $d=3$ - where we restrict ourselves henceforth - 
\begin{equation}
c_H = \sum_q C^2(q,\om=0) = \frac{y_i}{\sqrt{H}} \sum_{n=0}^\infty 
\left ( \frac{1}{a_i H}  + n + \frac{1}{2} \right )^{-3/2} \;\;,  \label{coope}
\end{equation}
is the contribution of the pair of Cooperons 
for a field $H$ along the $i$ space direction,
$a_i=4D_{i\perp}e\tau_{\phi}(T), \; 
y_i = 1/\{(4\pi)^3 N_F^2 \tau^4 \sqrt{e D_{i\perp}^3 D_i} \}$, and
$i\perp$ stands for the plane perpendicular to the axis $i$. 
Here we assume that $\tau_{\phi}^{-1}(T)\gg \tau_{S}^{-1}$.
For materials which are isotropic in the plane ($\parallel$)
but anisotropic in the direction perpendicular to the plane ($\perp$),
the following relation holds (c.f. also \cite{aalk} for a similar result)
\be
\frac{a_\perp}{a_\parallel} = \Big(\frac{y_\perp} {y_\parallel}\Big)^{-2} =
\sqrt{\frac{D_\perp}{D_\parallel}} \;\;. \label{ayd}
\ee

Eqs. (\ref{sig}) and (\ref{ayd}) yield a very good fit to the GMR data, with 
$a_\perp=0.190$ T$^{-1}$, $a_\parallel=0.085$ T$^{-1}$, 
$\Gamma_{1H}(H_\perp=4.58$T$)=0.4524\;\sigma^2$ and $r\equiv 4\sigma^3/(\pi 
\epsilon_F \alpha) = 0.03 \; \sigma^2$.
We have assumed that $\epsilon_F \gg \sqrt{\sigma^2-y_\pm}$, thus using
a total of 2 parameters per curve.
 
The disorder in the Dy/Sc superlattices arises probably mostly from
the interfaces, with Dy ($A_{Dy}=66$) providing 
spin-orbit scattering. 
Interface 
disorder (roughness) in superlattices amounts to effective
"bulk" disorder and anisotropic diffusion coefficients \cite{kc}.

	A big effective mass $m^*$ is required in order to fit this
theory with the data of ref.\cite{tuf}. This condition follows from the 
decay of the Cooperon for $H>H_{\phi}$, where the dephasing field 
$H_{\phi}\sim m^*/e\tau\epsilon_F\tau_{\phi}$\cite{aa}. 
Actually $m^* 
\geq 40 m$, $m$ being the free electron mass,
for hcp Sc near the Fermi surface and close to the points H and L
\cite{papa,siga}. 
However, GMR due to paramagnons can {\em in principle} appear for any $m^*$.

	The {\em sine qua non} condition for paramagnon-induced GMR is the
closeness of the constant $\Phi$ of the material in question to $\Phi_o$.
As in the limit of high electronic density the RPA interaction is appropriate 
we ought to compare the corresponding renormalized values $\Phi_{eff}$
with $\Phi_o$, {\em not} the 
bare $\Phi$ itself. 
In the $q \rightarrow 0$ static limit 
this procedure leads to the substitution
\begin{equation}
\Phi \rightarrow \Phi_{eff} = \frac{\Phi}{1-\Phi^2} \;\;.
\end{equation}
The bulk constant $\Phi$ has already been calculated within 
a band structure scheme
for a number of elements
by  Sigalas and Papaconstantopoulos\cite{siga}. Inserting $\Phi$ for hcp Sc in
$\Phi_{eff}(\Phi)$ yields a difference of 4.6\% from $\Phi_o=0.845$. 
Hence it is consistent to 
anticipate gigantic conductivity corrections for Sc and Sc based materials.

	This same sort of GMR has 
been seen in 
Sc films with disorder and also Er/Sc and Dy/Sc/Y superlattices
\cite{tsui}, i.e. only materials containing Sc. One may hypothesize
that the very high density of states of Sc around the Fermi level\cite{papa}
make Sc the dominant element and it is appropriate to use the value of $\Phi$ 
calculated for bulk Sc, as it would not change much for a Sc 
based superlattice. GMR is {\em not} seen in Y films, or Y based superlattices
\cite{tsui}, although Y, lying in the same column of the periodic table
as Sc just one row below it, has a very similar band structure to Sc. 
Nevertheless here the difference of $\Phi_{eff}$ from $\Phi_{o}$ 
is -14\%, and large transport corrections cannot arise.

	With regards to further predictions (or retrodictions) now,
$\Phi_{eff}$ of (fcc) Pt differs from $\Phi_{o}$ by 6.9\% and 
$\Phi_{eff}$ of (bcc) Rh differs by 0.4\% (they both have
high $m^*$'s and $N_F$ as well).
In an experiment done with polycrystalline samples of Pt and Rh
by Schulze\cite{schu} in 1941,
positive GMR was seen for both materials. It is likely that this was
the first signature of the paramagnon GMR seen experimentally.
Further, $\Phi_{eff}$ for (fcc) Cr differs by 4.5\% from $\Phi_{o}$.
More, the high-$T_c$ superconductor  Tl$_2$Ba$_2$CuO$_{6+\delta}$ (Tl-2201)
exhibits this sort of positive GMR \cite{boeb}, and in all appearances
the quasi 2-$d$ metal $\alpha$-(BEDT-TTF)$_2$KHg(SCN)$_4$ \cite{mcken}. Apparently the
value of $\Phi$ for these materials is not known at present.

	Our discussion so far presupposes that the microscopically 
calculated bare value of $\Phi$\cite{siga} is not really renormalized 
by paramagnons,
disorder etc. - {\em except} $\Phi_{eff}$ of course. Hertz et al.\cite{hertz} 
have shown that
there is no Migdal's theorem for paramagnons.
In that case, the first and second order vertex corrections 
are of the same order of magnitude as the bare paramagnon vertex, and 
presumably this is so for higher vertex corrections. However already the 
2nd order correction comes with a minus sign, and it is possible that 
a converging power series is thus formed for the total paramagnon vertex,
yielding a result close to the bare value. 
Also, it is not unlikely that the self-energy
and vertex corrections cancel each-other, as far as the $\Phi$ renormalization
is concerned, with the proviso above for $\Phi_{eff}$, in a manner analogous
to ref.\cite{ccm}.

	A few more comments about the ramifications of this theory are in
order. 
The correction to the density of states is 
small.
The correction to the Hall coefficient $R_H = \rho_{xy}/H$, with 
$\rho_{xy} = \sigma_{xy}/(\sigma_{xx}^2+\sigma_{xy}^2)$,
is usually given with the assumption that ${\Delta R/R}$ is small,
as e.g. in the work of Houghton et al.\cite{houg}. 
Here this is {\em not} the case 
and, assumming the cyclotron energy $\omega_c\ll\tau^{-1}$, we obtain
\begin{equation}
\frac {\delta R_H}{R_H} =  \frac{1 + \delta_{xy} - (1 + \delta_{xx})^2}
{(1 + \delta_{xx})^2} \;\;,
\delta_{ij}= \frac{\sigma_{Pij}}{\sigma_{oij}} \;\;,
\end{equation}
where $\sigma_{oij}$ is the usual Drude term and $\sigma_{Pij}$ the 
paramagnon contribution discussed above.
Probably this explains the gigantic correction to the Hall coefficient at
low temperature seen in ref. \cite{tuf}.

	In summary, we have shown that paramagnons in the weakly disordered 
regime can yield positive giant magnetoresistance
at low temperatures. The theory not only agrees with
experiment so far, but makes specific material-dependent predictions for 
future experiments as well. 

\begin{center} 
* * * 
\end{center}
The author is indebted to Yia-Chung Chang and Yuli Lyanda-Geller
for discussions and comments.
The author has enjoyed useful discussions with J. Betouras, E.N. Economou, 
A.J. Leggett, I. Martin, A.J. Millis,
D.A. Papaconstantopoulos,  M.M. Sigalas, F. Tsui and L. Xing.
This work was partially supported from NSF DMR 89-20538,
through MRL at UIUC.
\mbox{}

\newpage
{\bf Figure Captions}


\mbox{}

\noindent
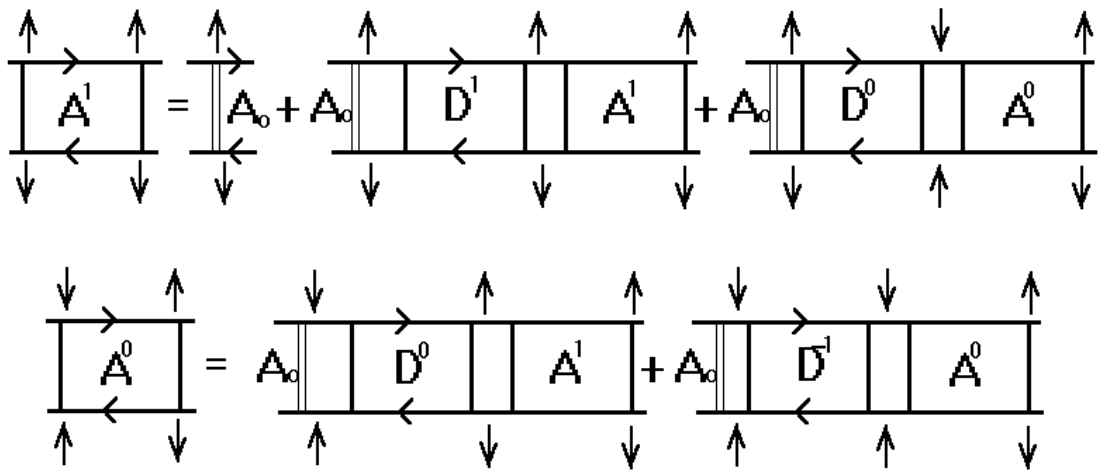
Fig. 1. The coupled Bethe-Salpeter equations obeyed by $A^i$.
Note the
explicit spin indices corresponding to the various parts of the diagrams.

\mbox{}

\noindent
Fig. 2. The basic diagrammatic blocks. The dashed lines in figs. b and c
denote impurity scattering.




\mbox{}

\noindent
Fig. 3. Plot of the giant magnetoresistance ${\Delta R/R} (0)$,
$\Delta R= R(H) - R(0)$. The points are the experimental data
of ref.\cite{tuf} for a typical Dy/Sc superlattice at $T=10^o$K and the lines
the theoretical fits, from eq. (\ref{sig}), with the constraint (\ref{ayd}).
The upper and lower lines correspond to the field $H$ being parallel and 
perpendicular to the superlattice growth axis.

\end{small}
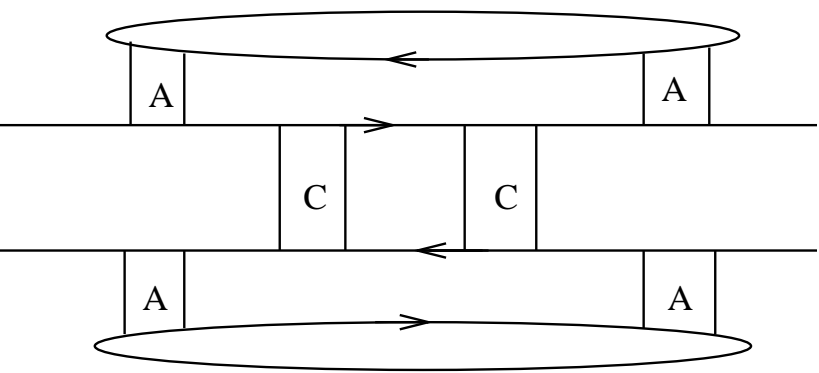
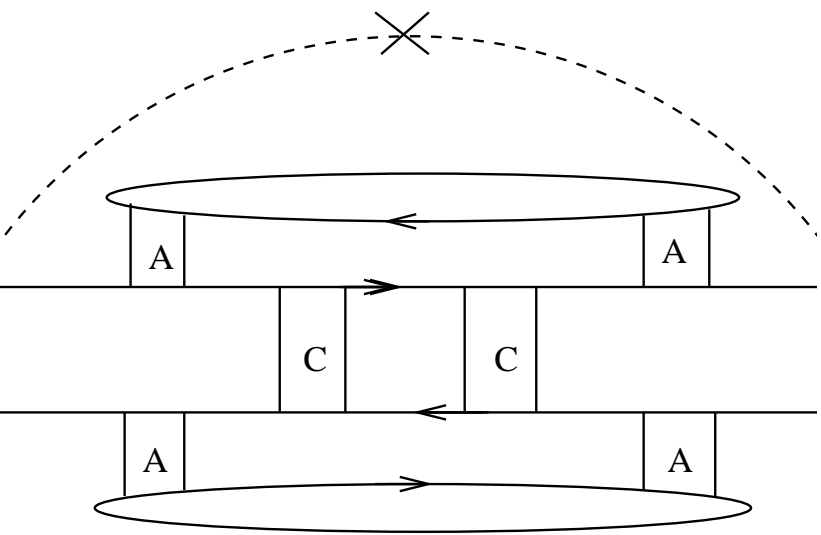
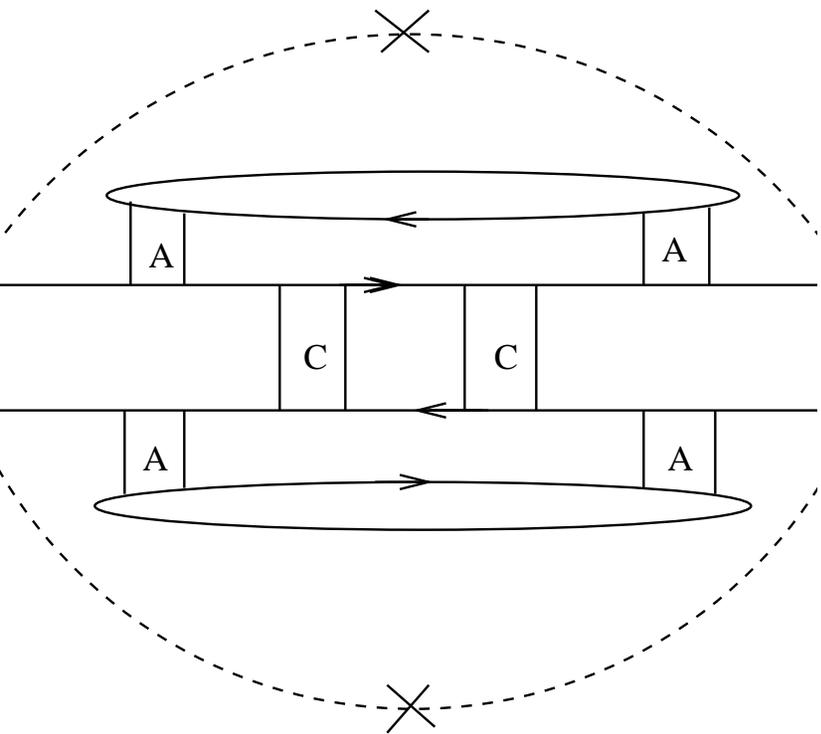
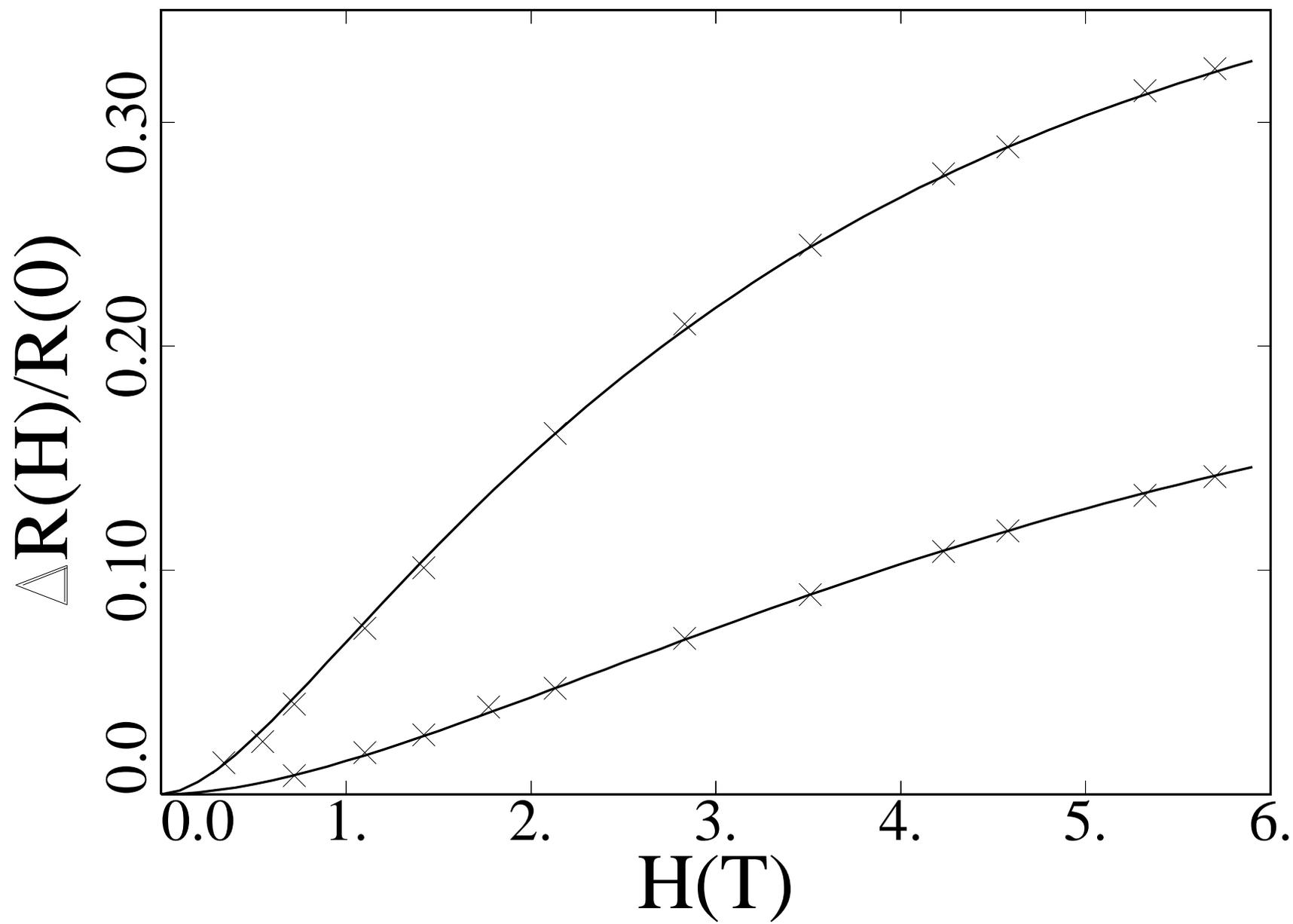
\end{document}